\documentclass[
    ]
    {aa}

\usepackage{graphicx}
\graphicspath{{figures/}}
\usepackage[varg]{txfonts}  % standard A&A font

\usepackage{commath}
\usepackage{siunitx}[=v2]
\usepackage[super]{nth}

\usepackage{xcolor}
\definecolor{click_color}{RGB}{46,48,118}

\usepackage{hyperref}
\hypersetup{
    pdffitwindow=true,
    pdfstartview={FitH},
    pdftitle={Waterfall PCA Folding},
    pdfauthor={Giampiero Naletto, Tomas Cassanelli},
    pdfcreator={Tomas Cassanelli},
    pdfproducer={Giampiero Naletto, Tomas Cassanelli},
    pdfnewwindow=true,
    colorlinks=true,
    linkcolor=click_color,
    citecolor=click_color,
    urlcolor=click_color
    }

\begin{document}

\title{A new technique to determine a pulsar period: the waterfall principal component analysis}

\author{
	T.\ Cassanelli\inst{\ref{inst:padova-physics},\ref{inst:uchile}}
	\and G.\ Naletto\inst{\ref{inst:padova-physics},\ref{inst:inaf}}
	\and G.\ Codogno\inst{\ref{inst:padova-engine}}
	\and C.\ Barbieri\inst{\ref{inst:padova-physics},\ref{inst:inaf}}
	\and E.\ Verroi\inst{\ref{inst:trento}}
	\and L.\ Zampieri\inst{\ref{inst:inaf}}
	}

\institute{
	University of Padova, Department of Physics and Astronomy ``Galileo Galilei'', Via Marzolo, 8, I-35131 Padova PD, Italy. \label{inst:padova-physics} \\
	\email{giampiero.naletto@unipd.it} 
	\and
	Department of Electrical Engineering, Universidad de Chile, Av. Tupper 2007, Santiago 8370451, Chile. \label{inst:uchile} \\
	\email{tcassanelli@ing.uchile.cl}
	\and
	INAF Astronomical Observatory of Padova, Vicolo dell'Osservatorio, 5, I-35122 Padova, Italy \label{inst:inaf}
	\and
	University of Padova, Department of Information Engineering, Via Gradenigo, 6/A, I-35131 Padova PD, Italy. \label{inst:padova-engine}
	\and
	Trento Institute for Fundamental Physics and Applications, Via Sommarive, 14 I-38123 Povo TN Italy. \label{inst:trento}
	}

% \date{Received ...; accepted ...}

	\abstract
    % CONTEXT (OPTIONAL)
	{}
    % AIMS (MANDATORY)
	{This paper describes a new technique for determining the optimal period of a pulsar and consequently its light curve.}
    % METHODS (MANDATORY)
	{The implemented technique makes use of the Principal Component Analysis (PCA) applied to the so-called waterfall diagram, which is a bidimensional representation of the pulsar acquired data. In this context we have developed the python package \texttt{pywpf} to easily retrieve the period with the presented method.}
    % RESULTS (MANDATORY)
	{We applied this technique to sets of data of the brightest pulsars in visible light that we obtained with the fast photon counter Iqueye. Our results are compared with those obtained by different and more classical analyses (e.g., epoch folding), showing that the periods so determined agree within the errors, and that the errors associated to the \textit{waterfall-PCA folding} technique are slightly smaller than those obtained by the $\chi^2$ epoch folding technique. We also simulated extremely noisy situations, showing that by means of a new merit function associated to the waterfall-PCA folding it is possible to get more confidence on the determined period with respect to the $\chi^2$ epoch folding technique.}
    % CONCLUSIONS (OPTIONAL)
	{}

	\keywords{
		Methods: data analysis --
		Techniques: miscellaneous --
		pulsars: light curve --
		Principal Component Analysis
		}

	\maketitle
 
\section{Introduction}
\label{sec:introduction}

High time resolution astrophysics (HTRA) investigates all types of celestial objects presenting rapid irradiance variability. Phenomena of interest include occultation measurements, oscillations in white dwarfs, flickering in cataclysmic variables, timing of pulsars, rapid variability in X-ray binaries and accreting compact objects. Among these, pulsars are perhaps the most studied HTRA objects, because of their exotic nature and the possibility to investigate fundamental relativistic physical and astrophysical problems. A peculiar characteristic of these objects is their (quasi) periodic signal, given by the combination of two factors: the beacon-like source, due to an approximately conical light emission beam of a very fast spinning neutron star, and the position of the Earth within the cone-scanned sky regions. With our instruments Aqueye/Aqueye+ \citep{2009JMOp...56..261B,2009ASSP....9..249B,2015SPIE.9504E..0CZ}, applied to the Asiago Copernicus telescope (Italy) and Iqueye \citep{2009A&A...508..531N}, applied to the ESO New Technology Telescope (La Silla, Chile), we have investigated the visible light emitting pulsars which are accessible with medium size telescopes \citep{2011AdSpR..47..365Z,2011MNRAS.412.2689G,2012A&A...548A..47G,2014MNRAS.439.2813Z,2019MNRAS.482..175S}. Only four of them can be seen with these telescopes, namely: PSR B0531+21 (Crab pulsar), PSR B0540-69 in the Large Magellanic Cloud, PSR B0833-45 (Vela pulsar), and the most recent PSR J1023+0038 \citep{2017NatAs...1..854A,2019MNRAS.485L.109Z}; another couple of weaker optical pulsars (PSR B0656+14, and PSR B0630+17, also known as Geminga) are accessible only with larger telescopes (for properties of optical pulsars, see e.g., \citealt{2011AdSpR..47.1281M}).

One of the most important analysis to be performed on the pulsars is the determination of their (quasi) periods, $P$. In addition, since these objects lose energy by particle/light emission, their angular momentum varies in time, and the pulsation period slowly increases.
Thus, from the analysis of how the period varies in time, information can be obtained on the pulsar lifetime ($\propto P/\dot{P}$) and evolution. To retrieve the most accurate value of the pulsar period from the measured data, and consequently to obtain the best light curves, different techniques have been developed. These techniques depend on the data that can be obtained with the available instrumentation. To have the best time resolution, the selected time bin has to be the smallest possible: but reducing the time bin also implies that the collected signal per time bin gets smaller and smaller, down to the limit of photon counting. So, depending on the instrumentation performance, generally there are two types of dataset format: if the time bin is relatively long with respect to the photon rate, practically all time bins have a non-zero value, and the signal is essentially a continuous function of time; instead, if the time bin is too short, many bins have a zero signal and a few have a \num{1} or few units value, providing a Poisson distribution signal.

\begin{figure}[t]
	\centering
	\includegraphics{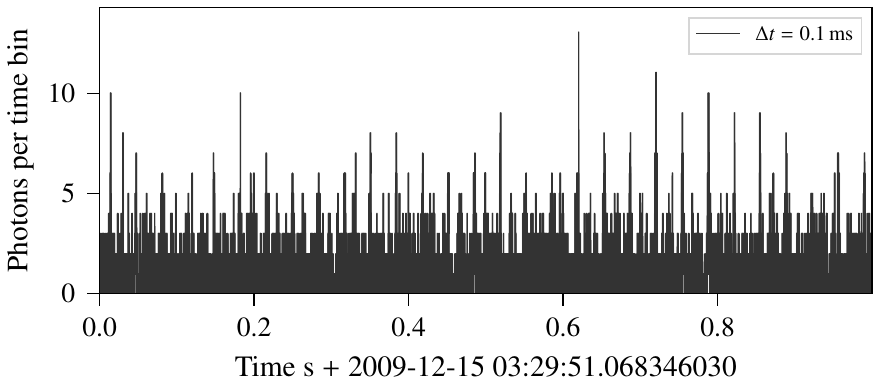}
	\caption{Example of light curve from the Crab pulsar (PSR B0531+21) acquired with Iqueye at NTT.
	Given the relatively low count rate, with a time bin of $\Delta t = \SI{0.1}{\milli\s}$ the periodic signal is not evident.}
	\label{fig:timestream_b0531+21}
\end{figure}

Generally, in these signals there is no clear evidence of a (quasi) periodicity (see for example Fig.~\ref{fig:timestream_b0531+21}), and dedicated analysis tools are requested to identify these features. In this paper we will briefly review the most common techniques to determine the pulsar period, and then we will describe how to get it by means of a novel technique which makes use of the Principal Component Analysis (PCA). 
For the latter we investigated two different situations: one, that we can define as ``low noise'', for which a very simple algorithm can be used; a second, the ``high noise'', for which a suitable merit function had to be defined. Indeed, the first situation applies to all the real cases we have investigated, that are the observations of the Crab pulsar taken in Asiago \citep{2012A&A...548A..47G}, PSR B0540-69 \citep{2011MNRAS.412.2689G} and the Vela pulsar \citep{2019MNRAS.482..175S} taken at the New Technology Telescope (NTT) in La Silla. The second situation has been reproduced artificially, by adding noise to the Vela pulsar data. The obtained results not only agree with those obtained by means of other well-established techniques, but the use of PCA provides a higher level of versatility which allows to get more confidence on the obtained results.

The structure of the paper is the following. Section \ref{sect:techniques} gives a short description of the standard techniques typically used to estimate a pulsar period, with some reminders also on the description of the so-called waterfall diagram. In Section \ref{sect:PCA} a description of the Principal Component Analysis is given. Then, Sections \ref{sect:PCA_Crab} and \ref{sect:PCA_pulsars} explain the simplest way to apply the PCA to a waterfall diagram to determine the pulsar period, which corresponds to the low noise case; the results obtained with this technique are also compared to those obtained by a more conventional technique. Finally, in Section \ref{sec:performance} there is a description of the method (waterfall-PCA folding) used in the highest noise case, in which a dedicated approach had to be considered; moreover, a comparative analysis with the results obtained by a standard technique is provided.
A summary of all the observations used for this analysis is reported in Table \ref{tab:observations}.

\begin{table}[t]
	\centering
	\caption{Observations used in the waterfall-PCA folding analysis. Acquisitions were performed with Iqueye instrument at the New Technology Telescope at La Silla Observatory, Chile. Arrival times of detected photons are corrected to the Solar System Barycenter (SSB) using the software \texttt{TEMPO2} \citep{2006ChJAS...6b.189H,2006MNRAS.372.1549E}.}
	\label{tab:observations}
	\begin{tabular}{llll}
		\hline
		Obs.~ID & PSR Source & Date & Total time \\ \hline
		OBS1 & B0540-69 & 2009-12-13 & \SI{3500}{\s} \\
		OBS2 & B0531+21 (Crab) & 2009-12-15 & \SI{7200}{\s} \\
		OBS3 & B0833-45 (Vela) & 2009-12-18 & \SI{3600}{\s} \\ \hline
	\end{tabular}
\end{table}

\section{Common techniques to determine the pulsar period}
\label{sect:techniques}

To analyze pulsar data, the first convenient step is to identify the possible presence of a signal periodicity simply using a Fourier Transform technique, for example the Fast Fourier Transform (FFT). This is a very simple and straightforward method, but has some intrinsic limitations: in some cases, periodic features are not so evident in an FFT, especially if the signal is rather noisy; the determination of the correct frequency is limited by the poor FFT frequency resolution, which depends on the number of input elements and on the ability of the software to deal with large arrays. In practice, because of these limitations, the FFT is used only for a preliminary identification of a possible periodic signal. In case the periodicity is confirmed, more accurate analyses have to be performed to improve the frequency measurements.

A very well-established technique to proceed in the data analysis, after having  identified the periodicity, is to \textit{fold} the signal, using the so-called \textit{epoch folding} technique \citep{1983ApJ...266..160L,1983ApJ...272..256L,1987A&A...180..275L,1992ApJ...398..146G,1996A&AS..117..197L}. In this case, a reasonable target period $P$ is assumed, either by the FFT analysis or by available information on the observed pulsar; then a trial period $P_\text{t}$ relatively close to $P$ is used for scanning a suitable time region around $P$ in search for the \textit{optimal period}, $P^*$. The search is realized by dividing $P_\text{t}$ into $N$ period time bins and co-adding the pulsar timing data modulo $P_\text{t}$ into these period time bins (NB: often, the \textit{phase} period is used instead of the \textit{time} period, with the phase period ranging between 0 and 1). By co-adding (folding) the signal over a large number of periods, assuming the period constant over the time of folding, the statistics per period time bin is largely increased, and the pulse shape is generated. 
Obviously, the more distant the trial period $P_\text{t}$ from the actual one $P^*$, the less accurate the light curve with respect to the correct one: to identify the best estimate of the actual period, and so to find the most accurate result, the standard procedure at this point is to vary the trial period by small amounts and to produce a set of light curves. By defining a suitable merit function, and iterating the process looking for the highest merit function value, it is possible to increase the accuracy of the trial period and finally to obtain the optimum solution.

Thus, the problem of finding the best light curve is finally reduced to the definition of the optimal merit function. A fairly common algorithm used to this aim is the calculation of the $\chi^2$ value for the estimated light curve \citep{1983ApJ...266..160L,1983ApJ...272..256L,1987A&A...180..275L,1996A&AS..117..197L}: indeed, it can be shown that the highest $\chi^2$ value provides the most accurate light curve. This approach is routinely adopted, and suitable algorithms can be found on dedicated software libraries (for example the timing analysis software for X-ray Astronomy \texttt{xronos}\footnote{\url{https://heasarc.gsfc.nasa.gov/docs/xanadu/xronos/xronos.html}} and \texttt{stingray}\footnote{\url{https://github.com/StingraySoftware/stingray}}).
Other adopted techniques are the {$Z$-test} \citep{1983A&A...128..245B} and the {$H$-test} \citep{1989A&A...221..180D,2010A&A...517L...9D}, which do not depend on the time binning.

Another accurate method, but not so frequently applicable, is the so-called waterfall diagram. Also with this method the signal is folded to increase the statistics, but the data analysis procedure is different. In this case, the total observation time is divided in $M$ time intervals, usually with $M \geq 20$, and folding is performed separately over each time interval with a common trial period $P_\text{t}$, so obtaining $M$ light curves. Then, each light curve is stacked as a row in a matrix,
\begin{equation*}
	\mathcal{W}=\del{w_{m, n}}\in\mathbb{R}^{M\times N};
\end{equation*}
by associating a color scale to the light curve signal intensity, this matrix can finally be represented as a color image.

\begin{figure*}[t]
	\centering
	\includegraphics{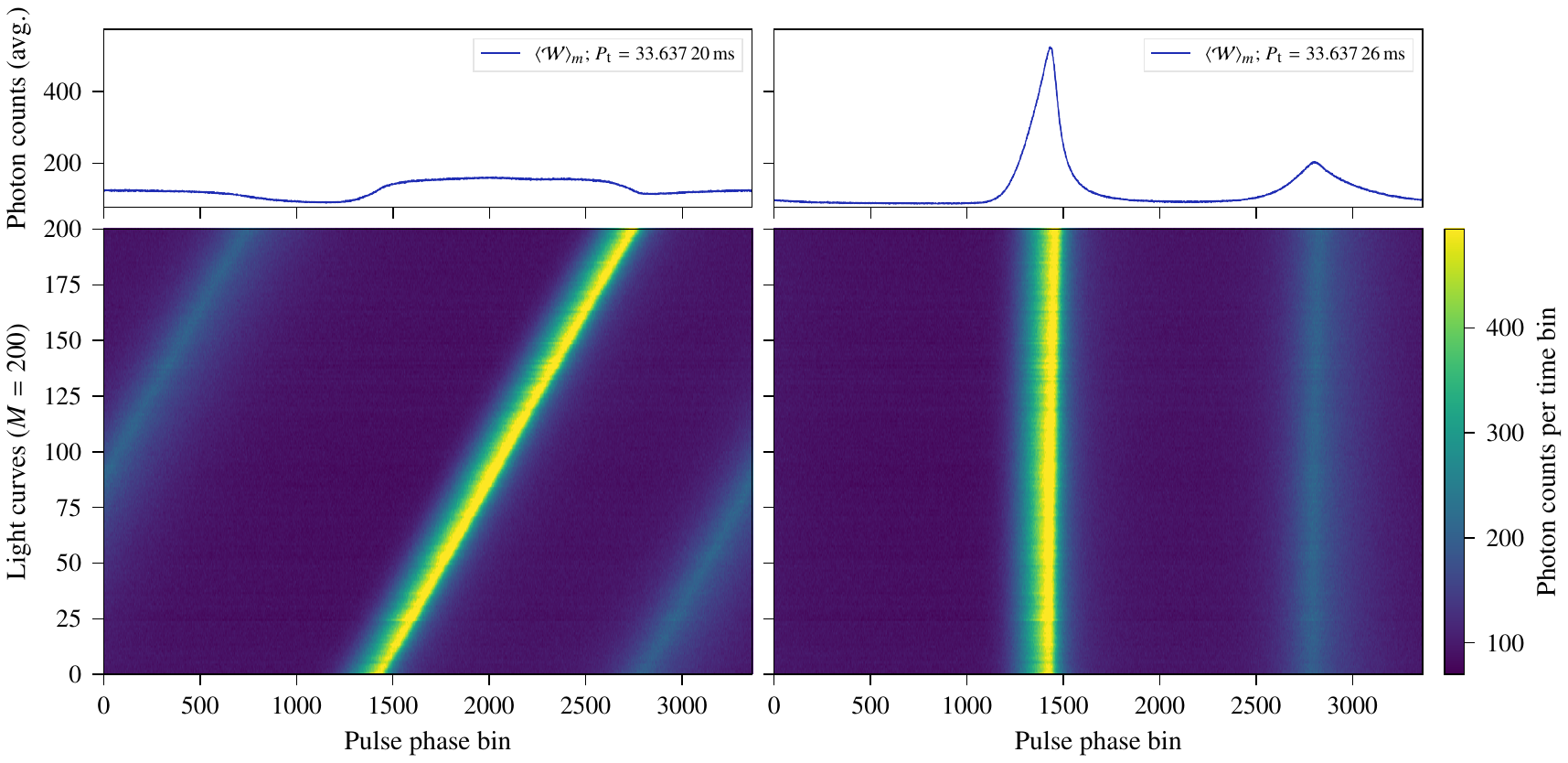}
	\caption{Waterfall diagram of a two hours Crab pulsar acquisition OBS2 (Table \ref{tab:observations}). The two columns have the same data sets, same number of divisions ($M=\num{200}$), and time bin ($\Delta t = \SI{10}{\micro\s}$); but different trial periods $P_\text{t}$. Left column has a $P_\text{t}=\SI{33.63720}{\milli\s}$ and the right column ha a $P_\text{t}=\SI{33.63726}{\milli\s}$, i.e., a \SI{60}{\nano\s} difference. Top panels are the average along the $m$-axis (number of divisions or number of light curves; $\left<\mathcal{W}\right>_m$) of each waterfall matrix $\mathcal{W}$. It is evident in this color representation that vertical straight lines appear only when the adopted period corresponds to the best estimation. In addition, a perfect straight line may not be possible since we are not considering pulsar spin-down or other effects in this calculation.}
	\label{fig:waterfall_single_b0531+21}
\end{figure*}

This method allows to detect, even by simple visual inspection, minute differences in the initial phase of the pulse. An example is shown in Fig.~\ref{fig:waterfall_single_b0531+21}, where the described method is applied with $M = \num{200}$ to the analysis of the period of the Crab pulsar, OBS2. 
On the left, the image obtained if the trial folding period $P_\text{t}$ differs from the actual one $P^*$: it is clearly evident that the initial phase of the pulse is changing because of the wrong folding period, producing tilted lines. Here we adopted a folding period $P_\text{t}$ largely different from $P^*$ to exaggerate the visual effect, but it can be shown that this method is very effective in measuring also very small inclinations of the lines corresponding to very small deviations of the folding period from the actual one.

In practice, as in the previous case, the procedure to follow is to introduce small variations of the trial period $P_\text{t}$ and to produce the corresponding waterfall diagrams; when the waterfall diagram shows straight vertical lines (see Fig.~\ref{fig:waterfall_single_b0531+21} right column), the optimal period has been obtained. Notwithstanding its simplicity, the accuracy of this method is quite remarkable, and period values as accurate as those obtained by the epoch folding technique can be obtained. An example of its accuracy is shown in Fig.~\ref{fig:waterfall_single_b0531+21}, where it can be seen, even if not so evident, a curvature of the ``vertical lines''. 
This waterfall diagram has been obtained from a \SI{\sim7200}{\s} long observation of the Crab pulsar, folding the data with a period which is correct at the middle of the observation and using $M = \num{200}$. 
This residual curvature is due to the phase shift occurred during the observing time as a consequence of the extremely small Crab pulsar regular spindown. This period variation can be measured with multiple observations, and we got, at the time of observation \citep{2014MNRAS.439.2813Z}, $\dot{P} \approx \SI{4.2e-13}{\s\per\s}$: this corresponds to a total change in the period of \SI{3}{\nano\s} from the beginning to the end of the observation, or equivalently to a total phase variation of \num{-0.0096}. These numbers demonstrate the goodness of this method in highlighting also these extremely small period/phase variations.

\begin{figure}[t]
	\centering
	\includegraphics{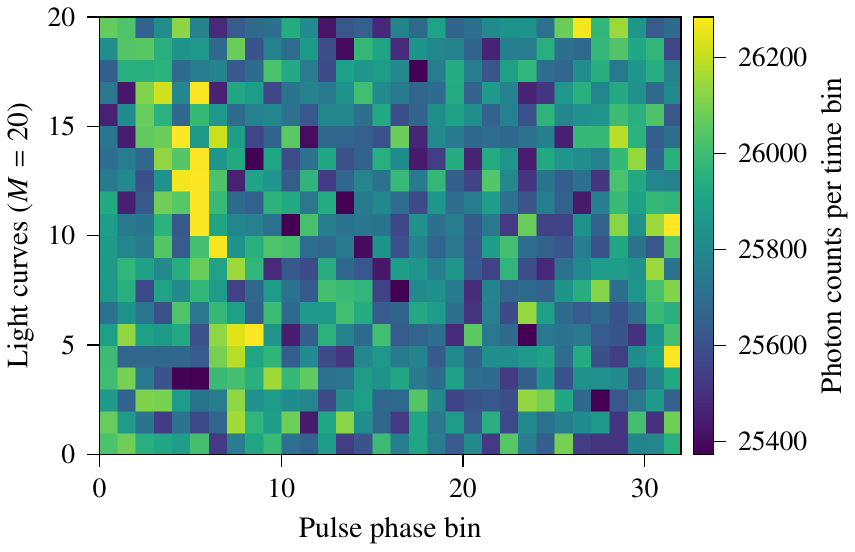}
	\caption{Waterfall diagram for the PSR B0833-45 (Vela pulsar) observed for \SI{180}{\min} with Iqueye at NTT (OBS3; Table \ref{tab:observations}). No vertical feature is evident, even if the data have been folded in just $M=\num{20}$ rows. The waterfall was generated with a $P_\text{t} = \SI{89.36715}{\milli\s}$ and a time bin of $\Delta t=\SI{2.793}{\milli\s}$.}
	\label{fig:waterfall_b0833-45}
\end{figure}

Unfortunately, this technique can be applied with good results only when the searched periodic signal is high with respect to the noise. When the noise is not negligible, the signal in the waterfall image has a very poor contrast making extremely difficult to detect any periodicity. As an example, Fig.~\ref{fig:waterfall_b0833-45} shows the waterfall diagram for the PSR B0833-45 (OBS3; Table \ref{tab:observations}) using $M = \num{20}$ and the nominal folding period: clearly, no straight vertical feature is evident, at least by eye.

The idea at the basis of this paper is to understand if, with a suitable software analysis, we can extract the ``hidden'' information on the actual pulsar period from waterfall images like the one shown in Fig.~\ref{fig:waterfall_b0833-45}, where the human eye or simple techniques fail to detect any feature. We show that the Principal Component Analysis, widely used in many scientific applications, offers such a tool.

\section{Principal Component Analysis}
\label{sect:PCA}

The Principal Components Analysis (PCA) (\citealt{doi:10.1080/14786440109462720}) is a mathematical tool that spans many fields of contemporary science: it is mainly used in image compression and more generally computer graphics, but also in statistics, mathematics, signal processing, meteorology, and so on. Depending on the field of application, this general technique is subject to small differences in implementation and so it is often named in different ways: in computer graphics and in statistics it is named as Principal Component Analysis; in signal processing it is better known as Karhunen-Lo\`{e}ve Transform (KLT; \citealt{Karhunen47,Loeve48}) and it is used for very low signal-to-noise ratio de-noising (in this field, while maintaining the basic characteristics of PCA, the algorithm and the method of application are different from those of PCA); other names with which this analysis technique is also known are Independent Component Analysis (ICA; \citealt{ICA01}), Hotelling Transform (HT; \citealt{Hotelling33,Hotelling36}), Proper Orthogonal Decomposition (POD; see for example \citealt{1993AnRFM..25..539B}), and others.

PCA is a relatively simple non-parametric method to extract useful information from data of otherwise difficult interpretation. Practically, PCA is a way of identifying patterns in data, and expressing the data in such a way as to highlight their similarities and differences: from a geometrical point of view, this corresponds to represent the data in a suitable reference frame such as to highlight their structure. For this, the algorithm uses an orthogonal transformation to convert a set of observations of possibly correlated variables into a set of values of uncorrelated variables called Principal Components (PCs). This transformation is defined in such a way that the first PC has as high a variance as possible (that is, it accounts for as much of the variability in the data as possible), and each following component has in turn the highest possible variance under the constraint that it has to be orthogonal to (uncorrelated with) the preceding components.

\begin{figure}[t]
	\centering
	% \pgfmathsetseed{2}
	% \begin{tikzpicture}
	% 	\edef\d{3}
	% 	\edef\r{0.07cm}
	% 	\coordinate (O) at (0, 0);
	% 	\coordinate (Or) at (\d/2, \d/2);

	% 	\draw[->, thick] (O) -- (\d, 0) node[below] {$x$};
	% 	\draw[->, thick] (O) -- (0, \d) node[left] {$y$};
		
	% 	\foreach \i in {1,2,...,40}{
	% 		\def\x{rand/2 + \d/2};
	% 		\def\y{rand/2 + \d/2};

	% 		\circrand{\y, \x}{\r};
	% 		\circrand{\y + .4, \x + .4}{\r};
	% 		\circrand{\y - .4, \x - .4}{\r};
	% 		% \circrand{rand + \d/3, (rand) + \d/4}{\r};
	% 		}

	% 	\draw[->, red, thick] (Or) -- ++ (-\d * .25, \d * .25) coordinate (V) node[left] {$\widehat{\mathbf{v}}$};
	% 	\draw[->, red, thick] (Or) -- ++ (\d * .4, \d * .4) coordinate (U) node[right] {$\widehat{\mathbf{u}}$};

	% \end{tikzpicture}
	\includegraphics{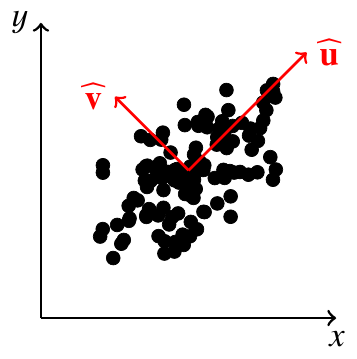}
	\caption{Given a bidimensional dataset with the distribution shown in this figure, the PCA allows to define the directions $\widehat{\mathbf{u}}$ and $\widehat{\mathbf{v}}$ of the orthogonal maximum variances, so providing the direction of largest variation.}
	\label{fig:pca}
\end{figure}

As a simple example, we can think of a two variable datasets represented in the $xy$ reference frame ($xy$-basis) as the set of points shown in Fig.~\ref{fig:pca}. It is evident in the figure that the data have a principal direction of variation, that has been identified with the $u$-axis; then, there is a second most important direction orthogonal to $u$, that is the $v$-axis. The PCA is a technique to identify these two directions, giving also them a priority on the basis of where the data have the largest dispersion. In practice, PCA allows to redefine the datasets in a new reference frame, $uv$-basis, having origin at the centroid of the data and oriented in such a way to get uncorrelated data, that is their co-variance with respect to the $(u, v)$ coordinates is zero. The directions of these axes are the so-called Principal Components.

The algorithm for redefining the data with respect to the PCs is rather simple, and consists of five steps. The first is to subtract the data mean value, which has to be done independently on the various variables: in the assumed two-dimensional data sample, the average value $\left<x\right>\del{\left<y\right>}$ has to be subtracted to all the $x$ ($y$) coordinates. Second, the covariance matrix $\mathcal{C}$ has to be calculated: in this example, $\mathcal{C}$ will be a $\num{2x2}$ matrix. The third step is the calculation of the eigenvectors $\mathbf{e}_\ell$ and the corresponding eigenvalues $\lambda_\ell$ of the covariance matrix. This process is equivalent to finding the reference frame in which the covariance matrix is diagonal, that is where the variables are uncorrelated: in this reference frame, the covariances (i.e., the non-diagonal elements of the matrix) are all equal to zero and only the variances (the diagonal elements) remain. The orientations of the axes of this reference frame are provided by the eigenvector set $\cbr{\mathbf{e}}$, which constitute an orthonormal basis, and the corresponding eigenvalue set $\cbr{\lambda}$ are equal to the variances along the directions of the corresponding eigenvectors $\mathbf{e}_\ell$. In the given example, the two eigenvectors provide the unitary direction vectors $\widehat{\mathbf{u}}$ and $\widehat{\mathbf{v}}$ shown in Fig.~\ref{fig:pca}; the corresponding eigenvalues are equal to the variances of the data along these two directions, with the eigenvalue associated to the $\widehat{\mathbf{u}}$ eigenvector being the largest.

In the fourth step the eigenvectors are ordered by the corresponding eigenvalue from highest to lowest, to put the direction given by the eigenvectors in order of importance. If the ordered eigenvectors are used as columns of a matrix, a so-called \textit{feature vector} $\mathbf{F}$ is obtained. In practice, this step allows to prioritize the PCs, with the most important ones retaining the largest amount of the data information. This is a very important step because it is possible to decide how much information of the original data to maintain, simply deciding how many PCs to use in the following step. If all the PCs are retained, clearly all the information is preserved. What often happens is that only the first ordered PCs have large variance, while the remaining ones provide a much smaller contribution: discarding the less important PCs, the complexity of the system can be largely reduced (``dimensional reduction'') at the well acceptable price of taking out only a minor amount of data information. With the final fifth step, the ``new'' dataset is obtained: the feature vector is used as a transformation matrix that takes the data points from the $xy$ reference frame to the $uv$ one by means of the equation
\begin{equation*}
	A(u,v) = \del{A(x,y) - \left<A(x,y)\right>}\mathbf{F},
\end{equation*}
where $A(x,y)$ is a point in the $xy$ reference frame, $\left<A(x,y)\right>$ is the dataset centroid, and $A(u, v)$ is the corresponding point in the $uv$ reference frame.

Routines for applying PCA algorithms can be found in many libraries of commercial scientific standard packages\footnote{For example in python scikit-learn: \url{https://scikit-learn.org}.}. Note that for the data reduction specifically required for waterfall-PCA folding routines, the software \texttt{pywpf}\footnote{The \texttt{pywpf} package: \url{https://github.com/tcassanelli/pywpf}, developed by T.~Cassanelli.} has been developed, and it has been used throughout the following sections.

\section{Application of PCA to the waterfall diagrams of the Crab pulsars}
\label{sect:PCA_Crab}

\begin{figure}[t]
	\centering
	% \begin{tikzpicture}[font=\small, x=(15:.5cm), y=(90:.5cm), z=(330:.5cm), >=stealth]
	% 	% \draw (0, 0, 0) -- (0, 0, 10) (4, 0, 0) -- (4, 0, 10);
	% 	\foreach \z in {0, 2, 4} 
	% 		\foreach \x in {0,...,8}
	% 			\foreach \y [evaluate={\b=random(0, 98);}] in {0,...,4}
	% 	    		\filldraw[viridis=\b] (\x, \y, \z) -- (\x+1, \y, \z) -- (\x+1, \y+1, \z) -- (\x, \y+1, \z) -- cycle (\x+.5, \y+.5, \z) node [yslant=tan(15)] {\b};

	% 	\draw[dashed] (0, 5, 0) -- (0, 5, 4);
	% 	\draw[dashed] (0, 0, 0) -- (0, 0, 4);
	% 	\draw[dashed] (9, 5, 0) -- (9, 5, 4);

	% 	\draw[->] (-1, 0, 0) -- node[yslant=tan(15), midway, fill=white] {$L$} (-1, 0, 4);
	% 	\draw[->] (1, -1, 4) -- node[yslant=tan(15), midway, fill=white] {$N$} (10, -1, 4);
	% 	\draw[->] (9, 1, 5) -- node[yslant=tan(15), midway, fill=white] {$M$} (9, 6, 5);

	% \end{tikzpicture}
	\includegraphics{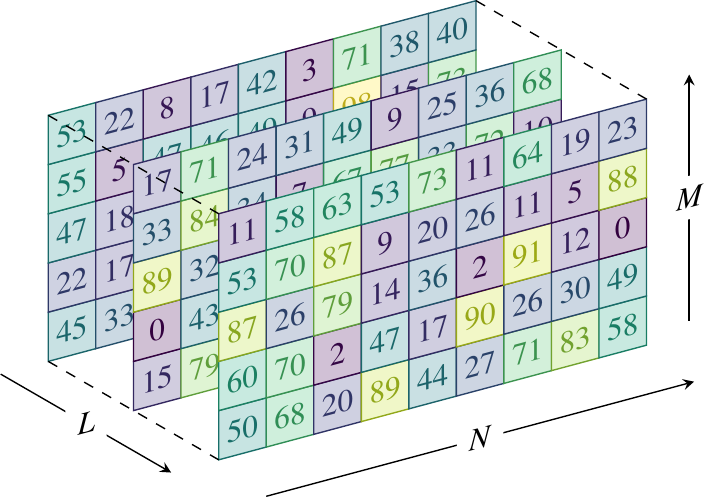}
	\caption{Waterfall diagrams, $\mathcal{W}_\ell$, where each sheet has $M\times N$ dimension. $M$ corresponds to the number of divisions (number of rows), $N$ is the number given by the trial period $P_\ell$ (viz., $P_\text{t}$ of the iteration $\ell$) divided by a chosen time bin $\Delta t$: $N = \text{round}\sbr{P_\ell/\Delta t}$. A data point has an amplitude given by the number of counts.
	$L$ corresponds to the total number of trial periods $P_\ell$ or number of iterations. For this particular case: $L=3, M=5$, and $N=9$. The PCA is computed for every $\mathcal{W}_\ell$, which returns \del{\ell,m} eigenvalues and eigenvectors.}
	\label{fig:wpca_matrix}
\end{figure}

To verify the possibility of applying the PCA technique to the waterfall diagrams to determine the optimal folding period, some tests have been done in the simplest case of the Crab pulsar (OBS2; see Table \ref{tab:observations}). The procedure we followed was first to produce different $M \times N$ waterfall diagrams $\mathcal{W}_\ell$ (where $N = \num{336}$ number of columns, i.e., the number of time bins in the folding period, and $M = \num{200}$ number of rows, that is the number of segments in which the whole acquisition has been divided), and then vary the folding period $P_\ell$, where $\ell$ is the index associated to the trial period (and $L$ is the total number of iterations). Fig.~\ref{fig:wpca_matrix} shows a graphical representation of the waterfalls while iterating over $\ell$: $\mathcal{W}_\ell$. 
Then we applied the PCA to each of them, considering these waterfall images as $M$-dimensional datasets (i.e., a set of \num{336} \num{200}-dimensional hyper-vectors corresponding to the columns of the waterfall $\mathcal{W}_\ell$). From this analysis, a set of $M$ $N$-component eigenvectors $\mathbf{e}_{\ell,m}$ with $\ell = 1, \dotso, L$; $m = 1, \dotso, M$ and of the corresponding eigenvalues $\lambda_{\ell,m}$ were obtained per each waterfall $\mathcal{W}_\ell$. Looking at the eigenvalues obtained with the waterfall corresponding to the nominal period $P^*$, it could be noticed that $\lambda_{*,1}$, the one associated to the first PC eigenvalue was largely dominant over the others, providing a ratio
\begin{equation*}
	\Lambda_{*,1} = \frac{\lambda_{*,1}}{\sum_{m=2}^M\lambda_{*,m}}=\num{0.99636},
\end{equation*}
and that all the components of the corresponding eigenvector $\mathbf{e}_{*,1}$ had substantially the same value (\num[parse-numbers=false]{\sim1/\sqrt{200}}). This is an expected result: when the period is the optimal one, all the rows of the waterfall are ideally identical, and so each hyper-vector is equal to the unitary hyper-vector multiplied by a constant, that is the value of the signal intensity of the corresponding time bin. This means that in the \num{200}-dimension space, the optimal $M$-dimensional dataset lies along the hyper-diagonal: in practice, all the points are ``aligned'' along this privileged direction. Along this direction, the dataset has the largest variance, which is the value provided by the eigenvalue, while in all the other perpendicular directions, the variance is almost null.

% \tikzset{
%   viridis/.style = {
%     text = viridis#1!75!black,
%     draw = viridis#1!75!black,
%     fill = viridis#1!25!white,
%   },
% }

\begin{figure}[t]
	\centering
	\includegraphics{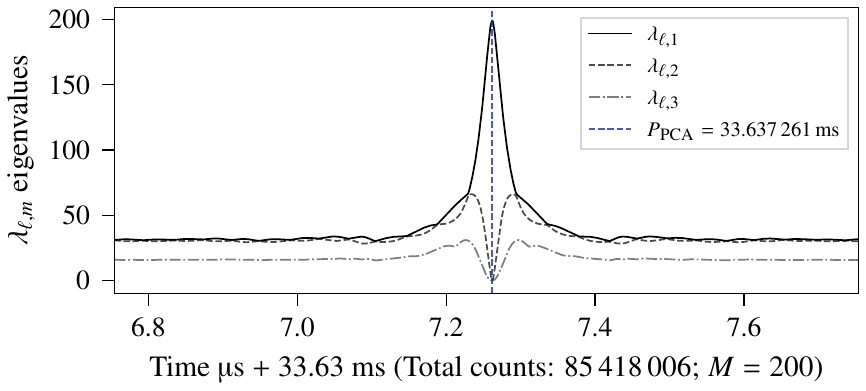}
	\caption{First PC eigenvalue, $\lambda_{\ell, 1}$, vs folding period ($\Delta_\text{s}=\SI{1}{\nano\s}$) for an acquisition of the Crab pulsar with Iqueye at NTT, OBS2 (Table \ref{tab:observations}). For this case $M=200$ has been chosen, and the optimum period found was $P_\text{PCA}=\SI{33.637261}{\milli\s}$ (with a search of $L=\num{1000}$ iterations). Second and third eigenvalues, $\lambda_{\ell, 2}$ and $\lambda_{\ell, 3}$, were also plotted. Notice that their amplitude is significantly lower than $\lambda_{\ell, 1}$, specially near $P_\text{PCA}$.}
	\label{fig:fe_b0531+21}
\end{figure}

By analyzing the eigenvalues for the various waterfalls $\mathcal{W}_\ell$, it could also be noticed that the ratio $\Lambda_{\ell,1}$ has a peak in correspondence of the optimal folding period. These results suggested to monitor the behavior of the eigenvalue of the first PC eigenvector to estimate the best folding period. In Fig.~\ref{fig:fe_b0531+21} the first PC eigenvalue, $\lambda_{\ell, 1}$, is shown. The folding period is changed multiple times and once the peak is found a smaller step can be defined, i.e., $\Delta_\text{s}=\SI{1}{\nano\s}$ for Fig.~\ref{fig:fe_b0531+21}.
In this case the highest value of the first PC eigenvalue is easily found in correspondence of a pulsar period of $P_\text{PCA}=\SI{33.637261}{\milli\s}$: compared to the one provided by the Jodrell Bank monthly ephemerides \citep{1993MNRAS.265.1003L}\footnote{\url{http://www.jb.man.ac.uk/~pulsar/crab.html}} at the same date of the Iqueye observations. This really remarkable result demonstrates the accuracy of this technique, at least for this simple case, definitely comparable if not better than other standard and much more well-established techniques at long radio wavelengths.

\begin{figure}[t]
	\centering
	\includegraphics{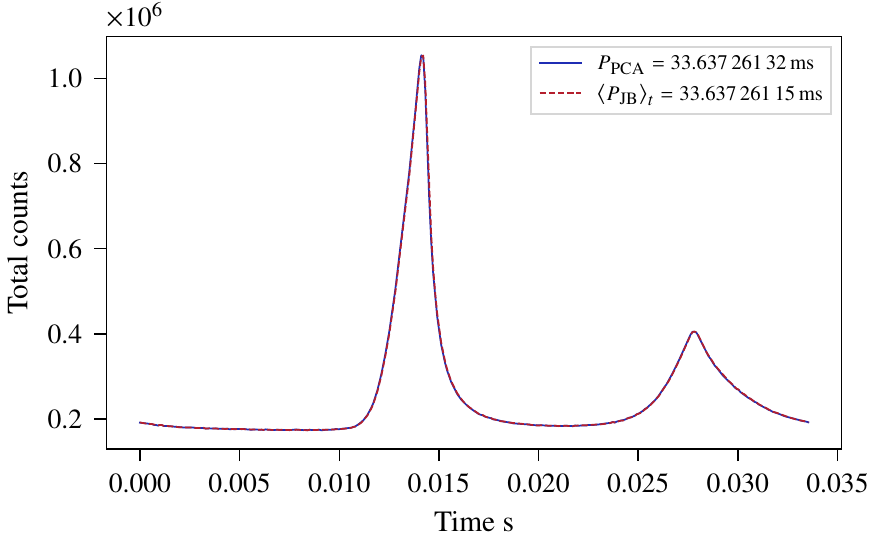}
	\caption{Crab pulsar (PSR B0531+21) light curve obtained by folding the data with the period found using the maximum value of the first PC eigenvalue, $P_\text{PCA}$ (as seen in Fig.~\ref{fig:fe_b0531+21}), and compared to the period from the Jodrell Bank monthly ephemerides, $P_\text{JB}$. The Jodrell Bank period was computed using \texttt{PINT} at the observation time and the period $\left<P_\text{JB}\right>_t$ was obtained  averaging over the whole transit (of duration $t=\SI{7200}{\s}$). The difference between the two methods is $\envert{P_\text{PCA}-\left<P_\text{JB}\right>_t}=\SI{0.1766}{\nano\s}$.}
	\label{fig:folded_b0531+21}
\end{figure}

From the obtained results in Fig.~\ref{fig:fe_b0531+21}, we can now use the estimated period from the first PC eigenvalue and fold the dataset. In Fig.~\ref{fig:folded_b0531+21} we show the folded data when using the maximum value from the first PC eigenvalue and the one from the Jodrell Bank monthly ephemerides \citep[see][Chapter~7]{2012hpa..book.....L}. The Jodrell Bank period, $P_\text{JB}$, was computed using \texttt{PINT} \citep{2019ascl.soft02007L} referenced to SSB at the observed timestamps, then the average period over the whole recording was computed. Notice that the computed $P_\text{JB}$ takes into account the pulsar's: Dispersion Measure (DM), $P$, $\dot{P}$, and other parameters (at a fixed epoch) to predict the period at a certain timestamp $t$.

\begin{figure}[t]
	\centering
	\includegraphics{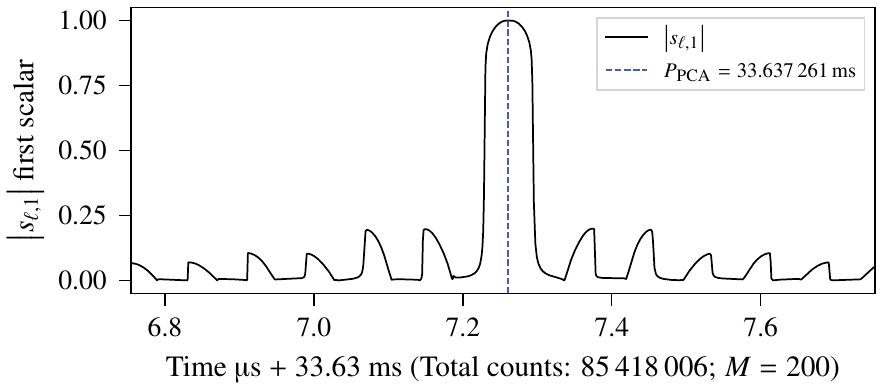}
	\caption{Absolute value of the scalar product, $\envert{s_{\ell, 1}}$ \eqref{eq:scalar}, of the first PC eigenvector with the \num{200}-dimension hyper-diagonal unitary vector as a function of the trial period in the case of the Crab pulsar. The scalar product (when high signal-to-noise ratio is available) reaches magnitude \num{1} in a plateau region. Eigenvalue and scalar are plotted together in Fig.~\ref{fig:merit_build_b0531+21}. The period $P_\text{PCA}$ has been computed with the first PC eigenvalue, Fig.~\ref{fig:fe_b0531+21}.}
	\label{fig:scalar_b0531+21}
\end{figure}

The application of the PCA to the waterfall diagrams allows to make a further check of the goodness of the obtained results. Since ideally at the optimal period $P^*$ the first PC eigenvector $\mathbf{e}_{*,1}$ should be parallel to the hyper-diagonal in the $M$-dimensional space, the optimal period can be obtained when the absolute value of the scalar product,
\begin{equation}
	s_{\ell, m} = {\mathbf{e}_{\ell,m}}\cdot\widehat{\mathbf{d}} \label{eq:scalar}
\end{equation}
of this eigenvector with the hyper-diagonal unitary vector $\widehat{\mathbf{d}}$ is maximum and ideally equal to \num{1}. The behavior of the scalar product $s_{\ell,1}$ in the previously examined case is shown in Fig.~\ref{fig:scalar_b0531+21}: this plot confirms the expected result, namely that the largest scalar product corresponds to the nominal period. Furthermore, the value of the scalar product assumes values close to \num{1} only in a small range of the abscissa, with a discontinuity at about \SI[parse-numbers=false]{\pm120}{\nano\s} from the nominal period. In fact, outside this range, the data are no longer well aligned, and the first PC eigenvector is far from being parallel to the hyper-diagonal. As we will see in Section \ref{sec:performance}, this consideration can be used as a result quality check when the noise is much larger than in this case.

\section{Application of PCA to the waterfall diagrams of other visible pulsars}
\label{sect:PCA_pulsars}

The same technique has been applied to the more noisy cases of the other two optical pulsars we observed in 2009 with Iqueye at the New Technology Telescope (NTT) in La Silla (Chile): the PSR B0540-69 in the Large Magellanic Cloud \citep{2011MNRAS.412.2689G}, and the PSR B0833-45 in the Vela supernova remnants \citep{2019MNRAS.482..175S}. In these cases, to reduce the noise, it has been necessary to take $M \leq 20$ only.

\begin{figure}[t]
	\centering
	\includegraphics{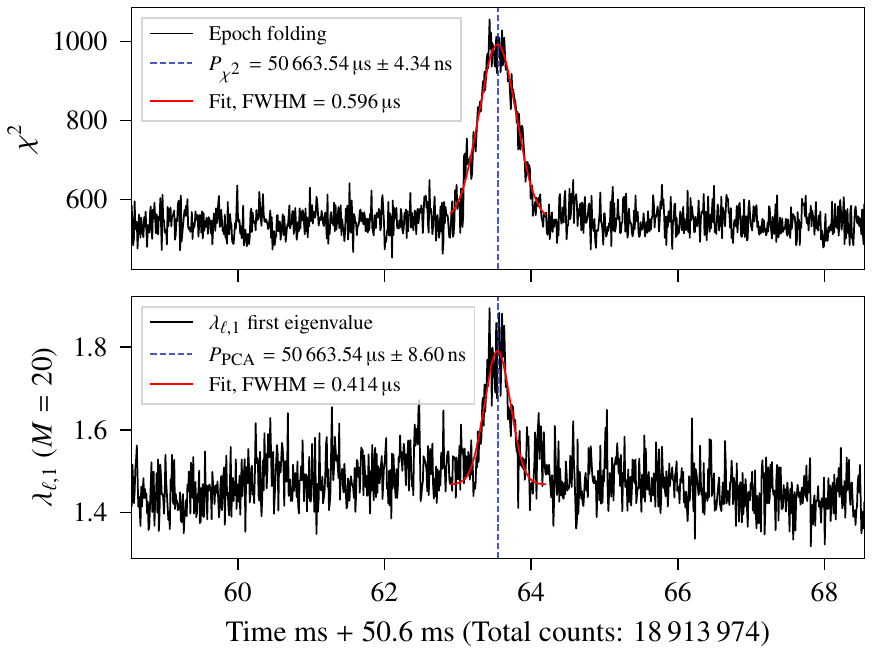}
	\caption{First eigenvalue compared to epoch folding for PSR B0540-69. Top panel shows $\chi^2$ optimization and bottom panel the first PC eigenvalue for a waterfall $\mathcal{W}$ with $M=20$ number of divisions. The eigenvalue, $\lambda_{\ell, 1}$ presents a noisier signal, i.e., less signal-to-noise ratio. However, the maximum from each plot reaches a similar result within less than \SI{10}{\nano\s}. The periods, $P_{\chi^2}$ and $P_\text{PCA}$, have been determined by a least-squares fit and adjusting a Gaussian (red line). The center of the Gaussian is the reported period and their errors are $\SI{4.34}{\nano\s}$ and $\SI{8.60}{\nano\s}$. The FWHM of the Gaussian is also reported.
	Both sets were started with the same initial conditions, $\Delta_\text{s}=\SI{10}{\nano\s}$, $\Delta{t}=\SI{1}{\milli\s}$, number of iterations, and initial trial period.}
	\label{fig:ef_fe_b0540-69}
\end{figure}

As shown in Fig.~\ref{fig:ef_fe_b0540-69} bottom panel, which refers to an acquisition of about \SI{58}{\minute} of the PSR B0540-69 (OBS1; see Table \ref{tab:observations}), the behavior of the first PC eigenvalue ($\lambda_{\ell,1}$) as a function of the trial period is rather noisy, and the peak is not so clearly defined as in the previous case. 
By means of a standard Gaussian fit, an improved estimated period can be found.
The obtained peak period corresponds to $P_\text{PCA}=\SI{50663540 +- 8.60}{\nano\s}$ (with the reported error from the Gaussian peak position \SI[parse-numbers=false]{\pm8.60}{\nano\s}). To cross check the goodness of this result, we determined the best period on the same dataset also by means of the more well-established epoch folding technique with $\chi^2$ optimum. The period obtained in this way is $P_{\chi^2} =\SI{50663540 +- 4.34}{\nano\s}$, Fig.~\ref{fig:ef_fe_b0540-69} top panel, where both periods agree within less than \SI{10}{\nano\s} (the time bin size $\Delta_\text{s}$).

From Fig.~\ref{fig:ef_fe_b0540-69} we can also see the Gaussian Full Width Half Maximum (FHWM), which differs by \SI{0.1}{\micro\s}. Nevertheless, from this simple estimated test, we can say that results from the epoch folding and the PCA, are very similar, but epoch folding shows a clear noise reduction, viz., it is superior.

\begin{figure}[t]
	\centering
	\includegraphics{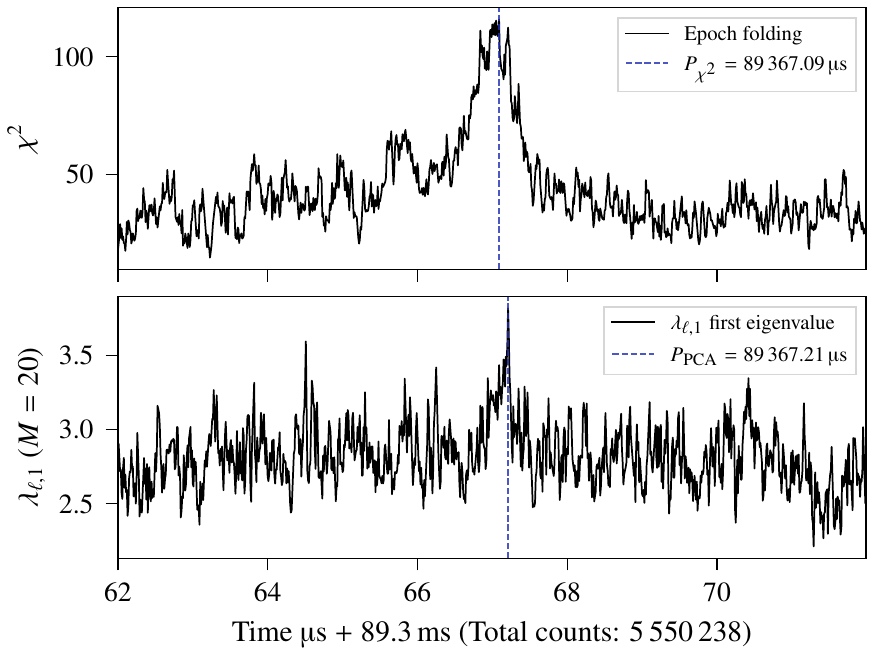}
	\caption{First eigenvalue compared to epoch folding for PSR B0833-45. Top panel shows $\chi^2$ optimization and bottom panel the first PC eigenvalue for a waterfall $\mathcal{W}$ with $M=20$ number of divisions. Similarly as in Fig.~\ref{fig:ef_fe_b0540-69}, the noise level is higher in the first PC. The optimum peaks are close, and their difference: $\Delta P = \envert{P_\text{PCA} - P_{\chi^2}}=\SI{120}{\nano\s}$. Both sets were started with the same initial conditions, $\Delta_\text{s}=\SI{10}{\nano\s}$, $\Delta{t}=\SI{2.793}{\milli\s}$, number of iterations, and initial trial period.}
	\label{fig:ef_fe_b0833-45}
\end{figure}

Finally, we also show here the results obtained in the case of the Vela pulsar (OBS3; Table \ref{tab:observations}). PSR B0540-69 is much less noisy than the Vela pulsar, which can be seen either in the epoch folding analysis or the first PC eigenvalue of Figs.~\ref{fig:ef_fe_b0540-69} and \ref{fig:ef_fe_b0833-45}.

In Fig.~\ref{fig:ef_fe_b0833-45} we have avoided to fit a Gaussian, since the signal is too noisy to retrieve a good measurement. However, the reported periods by only taking the maxima reach very similar values. The difference between the two periods is $\Delta P = \SI{120}{\nano\s}$.

\begin{figure*}
	\centering
	\includegraphics{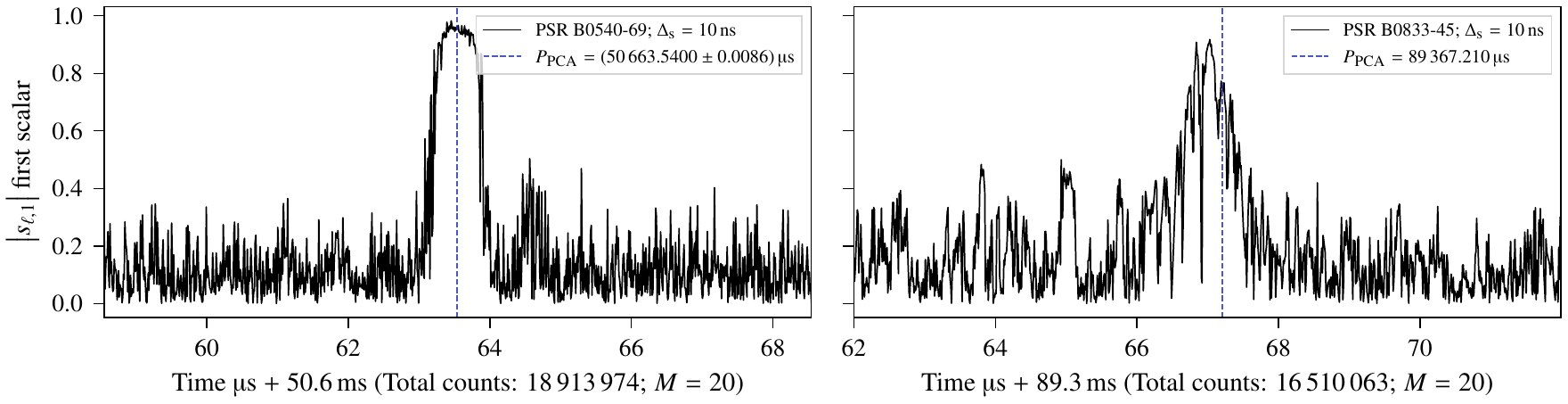}
	\caption{Absolute value of the scalar product, $\envert{s_{\ell, 1}}$, of the first PC eigenvector and the hyper-diagonal unit vector as a function of the trial period. Left: PSR B0540-69, same data as the one in Fig.~\ref{fig:ef_fe_b0540-69}. Right: PSR B0833-45, same data as the one in Fig.~\ref{fig:ef_fe_b0833-45}. It can be clearly seen that in both cases the optimum period lies within the plateau region, or where the maximum, \num{\sim1} is achieved. The dashed vertical line corresponds to $P_\text{PCA}$, the optimum value found by the first PC eigenvalue (from Figs.~\ref{fig:ef_fe_b0540-69} and \ref{fig:ef_fe_b0833-45}). This same behavior can be seen in the high signal-to-noise ratio case for PSR B0531+21 in Fig.~\ref{fig:scalar_b0531+21}. Total time and observation data are stated in Table \ref{tab:observations}.}
	\label{fig:s_b0540-69_b0833-45}
\end{figure*}

We monitored also in these two cases the behavior of the scalar product $s_{\ell,1}$ \eqref{eq:scalar} between the first PC eigenvector and the unity vector parallel to the hyper-diagonal in the $M$-dimensional space ($M = 20$ in this case). The obtained results are shown in Fig.~\ref{fig:s_b0540-69_b0833-45}. From these plots it is clear that the analysis of the scalar product $s_{\ell,1}$ is not as sensitive as the analysis of the first PC eigenvalue, $\lambda_{\ell, 1}$, to find the optimal period: while in Figs.~\ref{fig:ef_fe_b0540-69} and \ref{fig:ef_fe_b0833-45} there is a clear trend which allows to accurately define the highest value of the eigenvalue, in Fig.~\ref{fig:s_b0540-69_b0833-45} it is much more difficult to say where the maximum in the scalar product is. 
However, these plots allow to unequivocally say that the optimal period is located in the region where the absolute value of the scalar product is rather flat and reaches the highest values, larger than \num{0.8}: this is a sort of range of confidence which allows to say that the inferred period is within a plateau of flat region.
From Figs.~\ref{fig:ef_fe_b0540-69} and \ref{fig:ef_fe_b0833-45} we can also see that the same found period from the first PC eigenvalue, $P_\text{PCA}$, is within this high amplitude region in Fig.~\ref{fig:s_b0540-69_b0833-45}.

\section{Performance in case of a noise limited signal}
\label{sec:performance}

To properly check the possible performance of the application of PCA to the waterfalls to determine the pulsar periods, we realized a test to see what could be the limit for the detection of the optimal period when artificially increasing the noise; then we compared the results with those obtained applying epoch folding and $\chi^2$ optimization under the same conditions.

For this, we started with the same dataset used for the previous determination of the Vela pulsar period. This file collects the detection times of about \num{5.5} million photons and the obtained optimal period is $P^* = \SI{89367.09}{\micro\s}$ (top panel Fig.~\ref{fig:ef_fe_b0833-45}), as seen before. Producing the light curve with $N=\num{32}$ phase bins ($\Delta{t} \backsimeq \SI{2.793}{\milli\s}$), the average counts per bin value is \num{\sim173000}, and the peak-to-valley (PtV) of the Vela pulsar pulse profile is just less than \SI{1.6}{\percent} of it \citep{2019MNRAS.482..175S}.
Then we artificially increased the (white) noise simply adding photon counts uniformly distributed in time. For example, adding another \num{5.5} million counts corresponds to double the average with no change to the pulsar oscillation, and to bring the PtV to about \SI{0.8}{\percent}.
We analyzed several cases reducing the PtV value, and we stopped when we reached \num{23} million counts, which corresponds to a nominal PtV of less than \SI{0.4}{\percent}. It should be noticed that ideally the noise addition is not going to alter the pulsar light curve, so the expected determined optimal period should be noise independent.
To produce a correct comparison between the two methods, all period searches were performed using the same initial trial period $P_\text{t}=\SI{89.367}{\milli\s}$, and the time events have been binned at $\Delta t = \SI{2.793}{\milli\s}$. The time period variation used for the search was $\Delta_\text{s} = \SI{10}{\nano\s}$, which implies $L=\num{1000}$ iterations in a time range of \SI[parse-numbers=false]{\pm5}{\micro\s} around the nominal optimal period. 

\begin{figure}[t]
	\centering
	\includegraphics{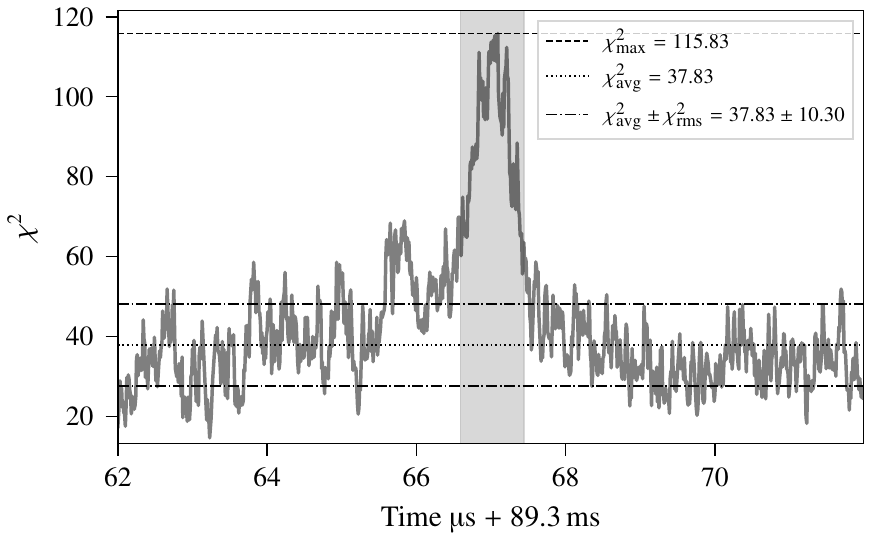}
	\caption{Example calculation of $\text{CP}_{\chi^2}$ value. Data corresponds to the first row in Table \ref{tab:CP} (original case for Vela pulsar; exactly same as in Fig.~\ref{fig:ef_fe_b0833-45} top panel). The average, $\chi^2_\text{avg}$, and rms, $\chi^2_\text{rms}$, are computed outside the gray area (off signal). The dataset achieves a $\text{CP}_{\chi^2}$ of \num{7.57}.}
	\label{fig:cp_build}
\end{figure}

\begin{figure}
	\centering
	\includegraphics{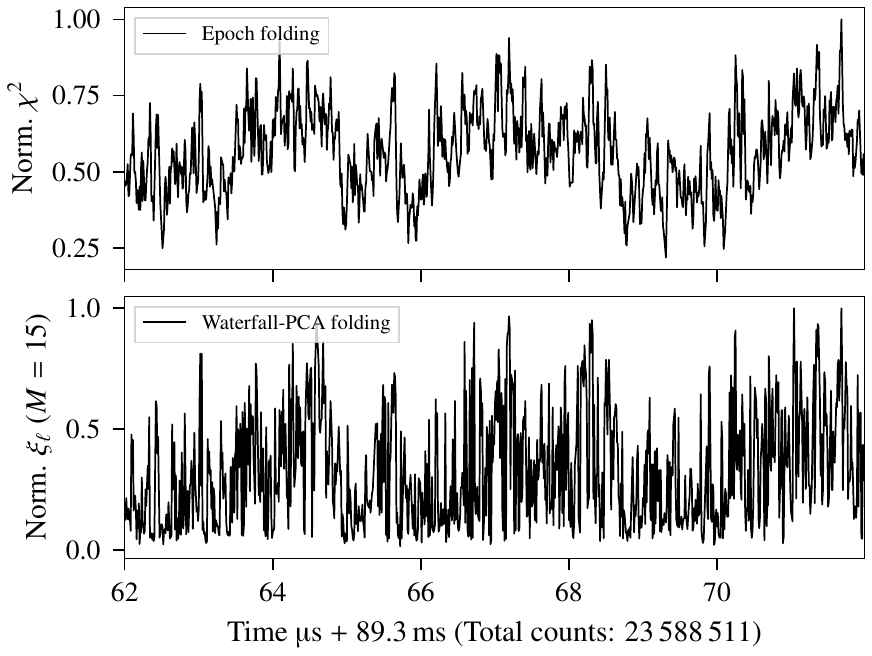}
	\caption{Epoch folding $\chi^2$ analysis and waterfall-PCA folding for searching the optimal period in the case of the maximum noise (\num{23} million total counts; bottom row Table \ref{tab:CP}). Both methods have been set with the same time bin ($\Delta{t}$), start trial period, and step period increase ($\Delta_\text{s}$). For comparison purposes both datasets have been normalized by their maxima. 
	% Fig.~\ref{fig:ef_wpf_b0833-45_n0} shows the original dataset with no noise added.
	}
	\label{fig:ef_wpf_b0833-45_n325}
\end{figure}

To quantify the goodness of the results when using the $\chi^2$ optimization, we adopted a \textit{confidence parameter} $\text{CP}_{\chi^2}$, defined as
\begin{equation}
	\text{CP}_{\chi^2} = \frac{\chi^2_\text{max} - \chi^2_\text{avg}}{\chi^2_\text{rms}},
	\label{eq:cp_chi}
\end{equation}
where $\chi^2_\text{max}$ is the maximum of $\chi^2$, $\chi^2_\text{avg}$ and $\chi^2_\text{rms}$ are the average and the standard deviation of $\chi^2$ outside the region of the peak.
The quantity $\text{CP}_{\chi^2}$ is substantially the signal-to-noise ratio for the $\chi^2$ function, but we prefer to call it in a different way because we consider as ``signal'' the pulsar light curve on top of the background ``noise'', and with this definition we avoid possible misunderstandings. A description on how to build the $\text{CP}_{\chi^2}$ is shown in Fig.~\ref{fig:cp_build}. We consider reliable the found period when $\text{CP}_{\chi^2} \gtrsim 5$.
In Table \ref{tab:CP}, second column, we report the $\text{CP}_{\chi^2}$ value for some of the cases we analyzed. We can confirm the robustness of this technique, since in all the cases the procedure returned a period corresponding to the nominal one within the errors. However, as clearly evident from the top panel of Fig.~\ref{fig:ef_wpf_b0833-45_n325} corresponding to the $\chi^2$ values for the most noisy case, we can say that the procedure pin-pointed the \textit{right value} in this extreme case essentially by chance: in fact, the maximum of this function is a single isolated value above the average in the region of the correct period, but its significance is very poor ($\text{CP}_{\chi^2} \approx\numrange{2}{3}$).

When we performed the analysis on these noisy datasets looking for the highest value of the first PC eigenvalue $\lambda_{\ell,1}$, we realized that even if this technique had been a powerful tool for determining the periods for all the visible pulsars, its performance under these extreme conditions was not as good as the $\chi^2$ optimization one: in fact, when we reached a value of about \numrange{12}{13} million total counts, the peak of $\lambda_{\ell,1}$ was no longer discernible from the noise. However, we also realized the possibility of improving the algorithm performance by using the additional available piece of information, i.e.\ the scalar product of the eigenvectors by the hyper-diagonal. In fact, for what said before, the optimal period is  associated to the maximum of the absolute value of the scalar product $s_{\ell,m}$ in \eqref{eq:scalar}, thus we can use it for defining a new and very accurate merit function.

\begin{figure}[t]
	\centering
	\includegraphics{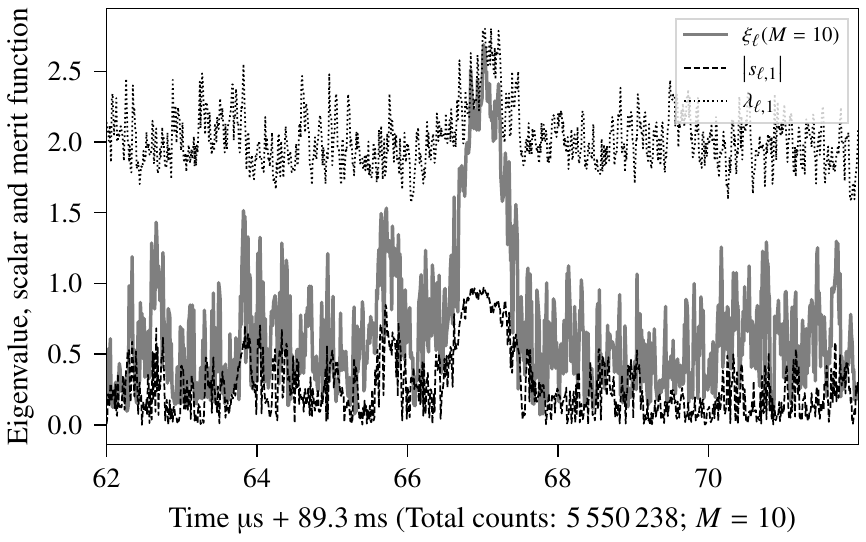}
	\caption{Construction of the waterfall-PCA folding merit function $\xi_\ell$ \eqref{eq:merit_function} for $M=10$ for the Vela pulsar observation (no noise added). The plot shows three curves, the maximum scalar \eqref{eq:scalar}, the eigenvalue corresponding to the same ${s_{\ell, \text{max}}}$, and the merit function $\xi_\ell$. The merit function shows a clear and stronger signal-to-noise ratio compared to the scalar or eigenvalues alone.}
	\label{fig:merit_build_b0833-45}
\end{figure}

\begin{figure}[t]
	\centering
	\includegraphics{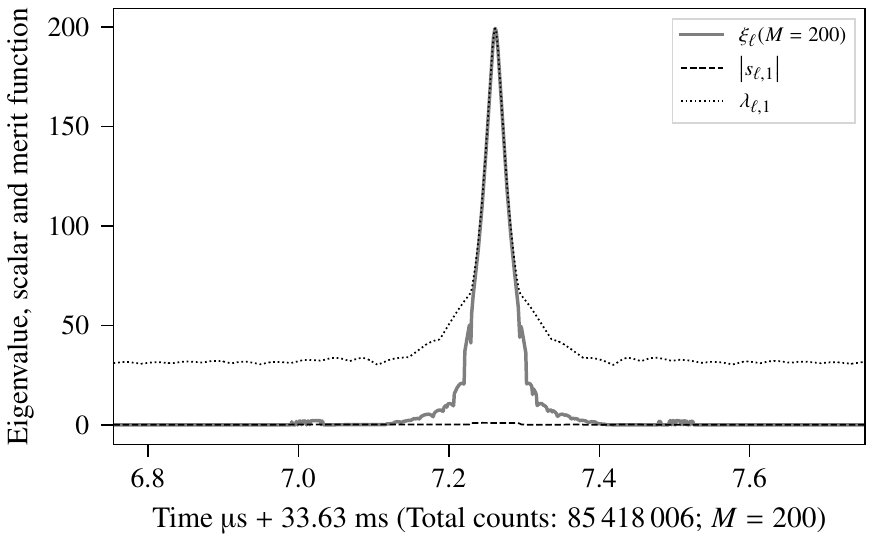}
	\caption{Construction of the waterfall-PCA folding merit function $\xi_\ell$ \eqref{eq:merit_function} for $M=200$ for the Crab pulsar observation. The plot shows three curves, the maximum scalar \eqref{eq:scalar}, the eigenvalue corresponding to the same ${s_{\ell, \text{max}}}$ (high signal-to-noise ratio case corresponds to the first PC eigenvalue), and the merit function $\xi_\ell$. It can be clearly seen the strength in signal-to-noise ratio for an ideal pulsar observation. The scalar, very low with respect to the rest (with \num{1} the maximum amplitude) of the curves is plotted in Fig.~\ref{fig:scalar_b0531+21}. Eigenvalues from the same dataset are shown in Fig.~\ref{fig:fe_b0531+21}.}
	\label{fig:merit_build_b0531+21}
\end{figure}

First, per each trial period $P_\ell$ and the relative waterfall $\mathcal{W}_\ell$, we consider the highest absolute value $s_{\ell,\text{max}}$ of all the scalar products
\begin{equation*}
	s_{\ell,m} = \mathbf{e}_{\ell,m}\cdot\widehat{\mathbf{d}} \quad\text{i.e.,}\quad s_{\ell,\text{max}} \equiv \max_{0<m\leq M} \envert{s_{\ell, m}}
\end{equation*}
(when the situation is getting very noisy, it may happen that the first PC eigenvalue is no longer associated to the highest absolute value of the scalar product, but to the second or the third; since in our analysis it is fundamental to find the eigenvector better co-aligned with the hyper-diagonal, we preferred to give priority to this parameter). Then, taken the  $\lambda_{\ell,\text{max}}$ eigenvalue, corresponding to the eigenvector $\mathbf{e}_{\ell,\text{max}}$ which provides the maximum scalar product $s_{\ell,\text{max}}$, we define
\begin{equation}
	\xi_\ell \equiv s_{\ell,\text{max}} \lambda_{\ell,\text{max}}
	\label{eq:merit_function}
\end{equation}
as the waterfall-PCA folding merit function in correspondence of the trial period $P_\ell$. Fig.~\ref{fig:merit_build_b0833-45} shows an example of this merit function with the Vela pulsar data, and Fig.~\ref{fig:merit_build_b0531+21} shows the same in the not noise-limited case of the Crab pulsar.
Finally, we use the $\xi$-function in the same way as the $\chi^2$ in the previous analysis: we calculate all the $\xi_\ell$ values as a function the trial periods $P_\ell$, getting the optimal period when $\xi_\ell$ is at its maximum value.

\begin{table}[t]
	\centering
	\caption{Values of the confidence parameters (CP) for some of the cases analyzed. In the second column there are the $\text{CP}_{\chi^2}$ values obtained with the $\chi^2$ optimization \eqref{eq:cp_chi}; in the third column there are $\text{CP}_\xi$ values obtained with the waterfall-PCA folding technique \eqref{eq:cp_xi}. Considerations to build the CP are shown graphically in Fig.~\ref{fig:cp_build}.}
	\label{tab:CP}
	\begin{tabular}{rcc}
	\hline
	Total counts & $\text{CP}_{\chi^2}$ & $\text{CP}_\xi$ \\ \hline
	\num{ 5550238} & \num[round-mode=places]{7.5719679700880613282} & \num[round-mode=places]{ 8.902653075876861384}\\
	\num{ 6937797} & \num[round-mode=places]{ 7.795348735544945562} & \num[round-mode=places]{ 7.783538251164919212}\\
	\num{ 8325357} & \num[round-mode=places]{ 6.545810578816168679} & \num[round-mode=places]{6.7878816608630613834}\\
	\num{11100476} & \num[round-mode=places]{4.9004322059073146534} & \num[round-mode=places]{4.9741310715937095315}\\
	\num{16650714} & \num[round-mode=places]{3.2242983415484940402} & \num[round-mode=places]{ 4.599785618375320961}\\
	\num{20813392} & \num[round-mode=places]{5.2498099184534838813} & \num[round-mode=places]{5.6631977182477574247}\\
	\num{22200952} & \num[round-mode=places]{2.2914981115258026279} & \num[round-mode=places]{4.2104143039728375957}\\
	\num{23588511} & \num[round-mode=places]{2.8863523397860828545} & \num[round-mode=places]{3.4999085464398565564} \\
	\hline
	\end{tabular}
\end{table}

\begin{figure}[t]
	\centering
	\includegraphics{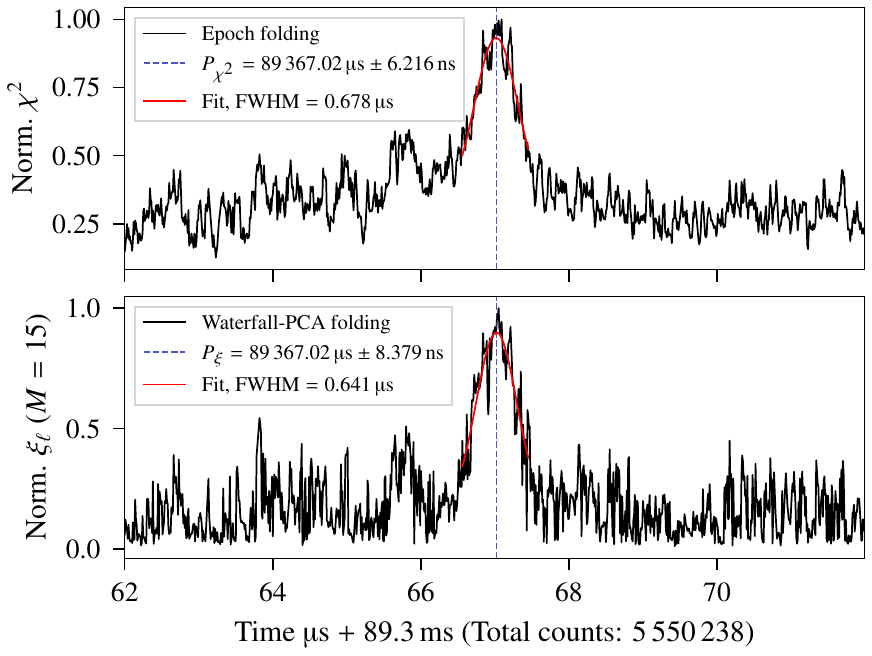}
	\caption{Comparison epoch folding and waterfall-PCA folding for the original Vela pulsar data (total counts: \num{5550238} and $\Delta_\text{s}=\SI{10}{\nano\s}$). For ease of comparison both outputs ($\chi^2$ and $\xi_\ell$) were normalized by their maxima. Red line shows the best Gaussian fit with an error in the centered position of $\SI{6.216}{\nano\s}$ and $\SI{8.379}{\nano\s}$, for $\chi^2$ and $\xi_\ell$.
	Fig.~\ref{fig:ef_wpf_b0833-45_n325} shows the same dataset when (white) noise is added.}
	\label{fig:ef_wpf_b0833-45_n0}
\end{figure}

As an example, in Fig.~\ref{fig:ef_wpf_b0833-45_n0} we compare the results of the optimal period Vela pulsar search using either $\chi^2$ or the just described $\xi$-function. The case $\xi_\ell$ $(M=15)$ shows a clear and strong central peak, as compared to the corresponding case shown in Fig.~\ref{fig:ef_fe_b0833-45} bottom panel, where only the first PC eigenvalue was used. The periods found using either the $\chi^2$ or the $\xi_\ell$ $(M=15)$ functions have a difference of the order of nanoseconds, well within the errors.

\begin{figure}[t]
	\centering
	\includegraphics{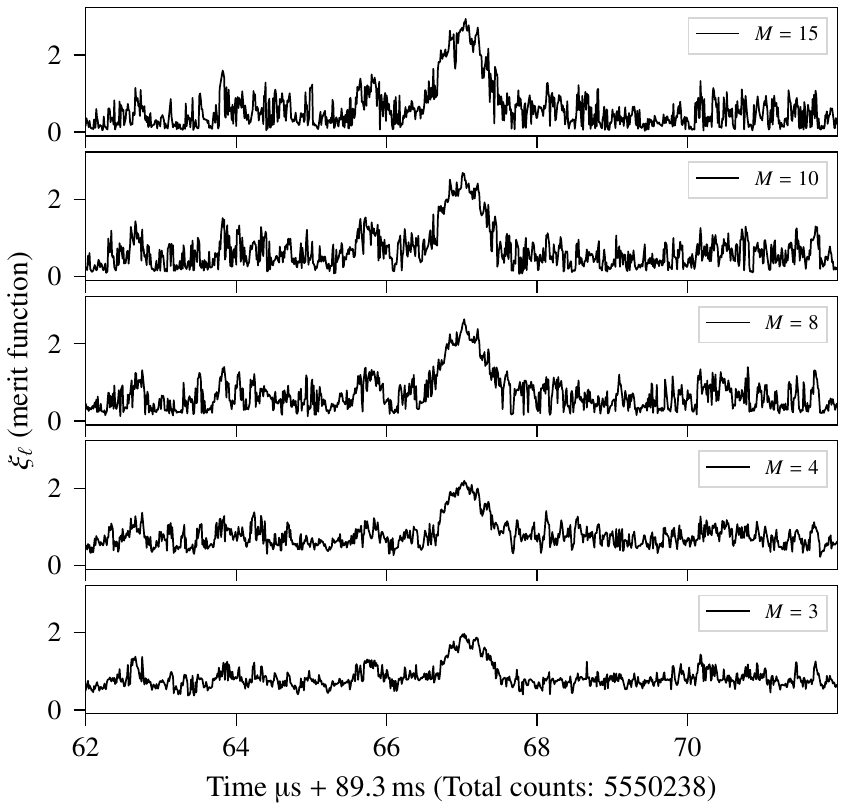}
	\caption{Waterfall-PCA folding merit function, $\xi_\ell$, for waterfalls $\mathcal{W}_\ell$ with different number of divisions, $M=3,4,8,10,$ and \num{15}. There is a clear improvement in the signal-to-noise ratio, while increasing $M$, up to a certain value (for this case for $M>10$ there is a slight degradation in the signal). The Vela pulsar data have the the original number of counts. The second row from top to bottom corresponds to the same merit function plotted in Fig.~\ref{fig:merit_build_b0833-45}.}
	\label{fig:merit_M}
\end{figure}

During our analysis, we noticed that using a different number $M$ of divisions slightly different results could be obtained (see for example Fig.~\ref{fig:merit_M}). Thus, we decided to provide a still more accurate merit function averaging the obtained $\xi_\ell$ for all the used values of $M$. For example, in this test we varied $M$ in the range $M = 3, 4, \dotso, 20$, obtaining\footnote{Note that this differs from summing along the $m$ index (introduced in Section \ref{sect:PCA_Crab}). The $m$ index does not exists in the merit function $\xi_\ell$, since its dependence vanishes in equation (\eqref{eq:merit_function}).}:
\begin{equation}
	\left<\xi_\ell\right>_M \equiv \frac{1}{M -3} \sum_{M=3}^{M=20}\del{\xi_\ell}_M.
	\label{eq:merit_mean}
\end{equation}

For quantifying the goodness of the results obtained by using \eqref{eq:merit_mean}, also in this case we defined a confidence parameter
\begin{equation}
	\text{CP}_\xi = \frac{{\left<\xi_\ell\right>_M}_\text{max} - {\left<\xi_\ell\right>_M}_\text{avg}}{{\left<\xi_\ell\right>_M}_\text{rms}}.
	\label{eq:cp_xi}
\end{equation}
where ${\left<\xi_\ell\right>_M}_\text{max}$ is the maximum of ${\left<\xi_\ell\right>_M}$, and ${\left<\xi_\ell\right>_M}_\text{avg}$ and ${\left<\xi_\ell\right>_M}_\text{rms}$ are the average and the standard deviation of ${\left<\xi_\ell\right>_M}$ outside the region of the peak. Looking at the corresponding expressions of \eqref{eq:cp_chi} it is obvious that we can use these two parameters for comparing the results of the two methods.

\begin{figure}[t]
	\centering
	\includegraphics{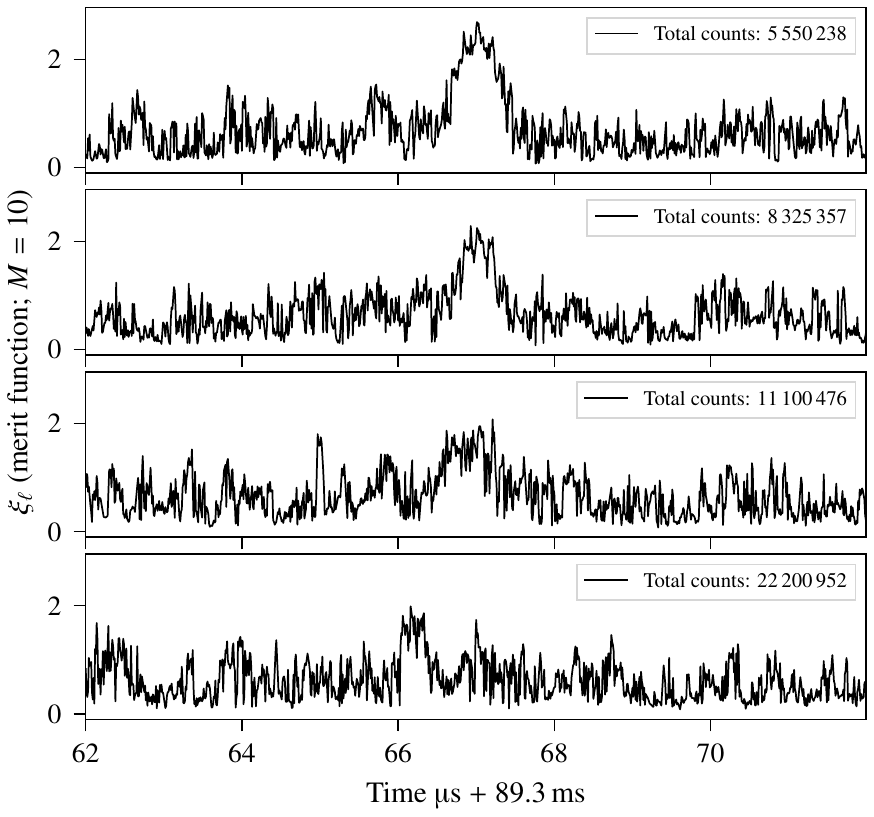}
	\caption{Waterfall-PCA folding merit function, $\xi_\ell$, and its robustness while adding (white) noise to the data. Top row corresponds to the original Vela pulsar data (same as in Fig.~\ref{fig:merit_build_b0833-45}). The noise increases from top to bottom. The number of divisions, $M=10$, was kept fixed for all calculations.}
	\label{fig:merit_noise}
\end{figure}

Table \ref{tab:CP}, third column, provides the values for the confidence parameters $\text{CP}_\xi$ determined in this case: by comparing these values with the corresponding $\text{CP}_{\chi^2}$ ones, we can see that the waterfall-PCA folding technique provides some more confidence about the goodness of the results. Figs.~\ref{fig:merit_M} and \ref{fig:merit_noise} show how the merit function changes with the number of division or when (white) noise is increased.

\begin{figure}[t]
	\centering
	\includegraphics{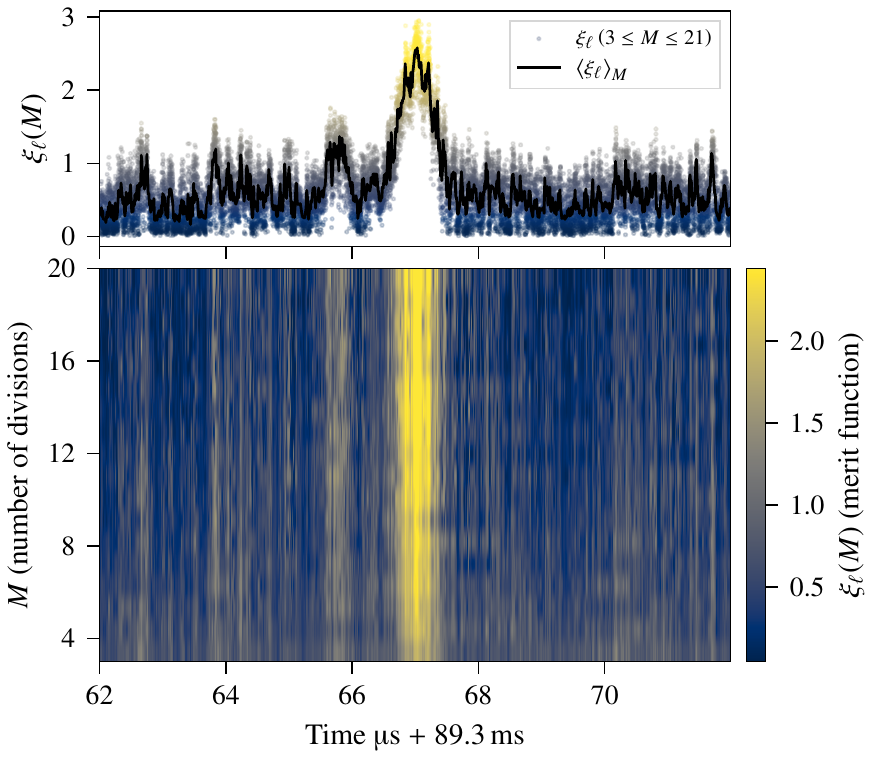}
	\caption{The color map rows show the amplitude of each waterfall-PCA folding merit function, $\xi_\ell(M)$. The top panel shows the average of all $M$ values, $3\leq M \leq 20$, and it is then plotted as a solid line. At the same time a scatter plot of the original data is over plotted (with same amplitude color bar). There is a clear noise reduction in the $\left<\xi_\ell\right>_M$ computed \eqref{eq:merit_mean}. The case corresponds to the nominal Vela pulsar of \num{5.5} million counts.
	The solid line, has a $CP_\xi=8.9$ (Table \ref{tab:CP}). Some discrete merit function (rows in the color map) are plotted in Fig.~\ref{fig:merit_M}. }
	\label{fig:merit_M_mean}
\end{figure}

A better way to visualize how the average over $M$ improves the waterfall-PCA folding function, $\left<\xi_\ell\right>_M$ is seen in Fig.~\ref{fig:merit_M_mean}. This merit function waterfall (different from the light curve waterfalls in Figs.~\ref{fig:waterfall_single_b0531+21}, and \ref{fig:waterfall_b0833-45}) shows $\xi_\ell=\xi_\ell(M)$, and the top panel takes the average overall $M$ cases \eqref{eq:merit_mean}. A clear noise reduction can be seen, since noise cancels out while the signal adds up. The computation of $\text{CP}_\xi$ (third row Table \ref{tab:CP}) is done only over the average, lower noise, case $\left<\xi_\ell\right>_M$.

To further confirm the obtained results, this analysis was performed three times adding (white) noise to the original signal with different random seeds. All the three cases showed equivalent results, i.e., equivalent numbers in Table \ref{tab:CP}.

Some considerations have to be done in relation to the analysis of the latter and more noisy case. This is somehow a limiting situation for both techniques (see Fig.~\ref{fig:ef_wpf_b0833-45_n325}). For the epoch folding $\chi^2$ technique, we verified not only that increasing the noise there is no longer convergence on the optimal period, but also that changing the time bin in this case occasionally provides the determination of a wrong period. For what concerns the waterfall-PCA folding technique, in \num{8} out the \num{20} different checked $M$ values, keeping fixed the time bin, the algorithm returned the same wrong peak as the $\chi^2$ one. However, because of the possibility of having $M-3$ datasets, we can be more confident about the goodness of the peak occurring with larger frequency. Moreover, because of this possible wrong detection, we have to specify that the corresponding value of $\text{CP}_\xi$ reported in Table \ref{tab:CP} has been calculated always considering the peak measured at the nominal period, independent of the possible presence of other higher peaks. As can be seen in Fig.~\ref{fig:ef_wpf_b0833-45_n325} and \ref{fig:merit_M}, the wrong detections are concentrated at the low $M$ values, indicating that this technique is more accurate for high $M$ values. 
The latter and noisiest case is a limiting situation for both techniques.
% In practice, with both techniques we are at the limit of the period detection, but the waterfall-PCA folding allows some more confidence on the goodness of the result because of the possibility of making some more statistical analysis (e.g., average over $M$, $\left<\xi_\ell\right>_M$; Fig.~\ref{fig:merit_M_mean}).

\subsection{Future considerations}
\label{sec:future_considerations}

% weighted PCA, know about on/off source statistics and weight data

Some additional points can be taken at the time to execute PCA over these types of datasets. So far we have assumed that there is no instrumental effect added to our data, which in principle is not true. By observing on and off pulse, a statistical information about the instrument itself can be extracted. This statistical information will be most sensitive when binning is applied to the data, which may change how the data is distributed.
Using this extra statistical information from the observation, a weighted-PCA algorithm \citep{2015MNRAS.446.3545D} can be implemented to further considering the photon statistics, and possibly improve the signal search. This is particularly important since the PCA method searches for the variance direction but does not distinguish whether the variance comes from signal or from noise.

% is this worth mentioning?
In addition, one could use information from pulsars, such as PSR B0531+21, as timing calibrators in a \textit{nodding} strategy. This is, observe PSR B0531+21 (calibrator), then observe a nearby target source, and then back again to the calibrator. After this we could examine the known drift in the rubidium clock and check discrepancies with the GPS pulse-per-second timing correction \citep{2009A&A...508..531N}. A precise timing of the arrival wavefront will increase the performance of either waterfall-PCA or epoch folding. This is specifically suitable at the time to search for optical signals of pulsars not previously seen at these wavelengths.

% perhaps some closing remarks here

\section{Conclusions}
\label{sec:conclusions}

\begin{enumerate}
	\item We have described a new technique for obtaining the light curves of fast periodic objects like pulsars using the Principal Component
	Analysis, viz., the waterfall-PCA folding. A complete and easy to use package has been developed for this purpose, \texttt{pywpf}.
	\item The obtained results have been compared to other more classical methods as the $\chi^2$ epoch folding technique. The performed analysis shows that the waterfall-PCA folding techniques works as expected and provides a small improvement against the classical method.
	The waterfall-PCA folding, in addition, allows to check the goodness of the obtained results, providing a simple tool for determining if the optimum period is within a range of confidence.
	
	% \item We conclude that this new technique is a rather powerful tool for performing this type of analysis. We applied it only to data acquired with our instrument Iqueye (Table \ref{tab:observations}), which works in visible light; but clearly it can be applied to all spectral ranges (e.g., X-rays).
	% We plan to apply it to even noisier signals, where also the epoch folding technique fails, to see if it is possible to better investigate periodicities in the light curves of extremely faint objects.
	% The waterfall PCA folding also checks the results' robustness, providing a simple tool to explore their confidence range.

	\item This new technique is a rather powerful tool for performing this type of analysis. We applied it only to data acquired with our instrument Iqueye (Table \ref{tab:observations}), which works in visible light. Although a specific implementation is beyond the purpose of this work, in principle this technique could be extended also to other spectral ranges (e.g., infrared, X-rays), if significant photon counting statistics are available. The required condition is that an adequate number of counts per period/bin can be collected in a rather limited interval of time (which is usually not the case for, e.g., the gamma-ray band), before the pulsar slow down or timing irregularities start to significantly affect the accuracy of the period determination. In the latter case, the evolution of the pulsar period has to be incorporated in the analysis and a timing solution is needed. When it can be fruitfully applied, the waterfall-PCA folding also checks the results' robustness, providing a simple tool to explore their confidence range.
\end{enumerate}

\begin{acknowledgements}
This work is based on observations made with ESO Telescopes at the La Silla Observatory under programme IDs 082.D-0382 and 084.D-0328(A), and on observations collected at the Copernico Telescope (Asiago, Italy) of the INAF-Osservatorio Astronomico di Padova.

We acknowledge the use of the Crab pulsar radio ephemerides available at the web site of the Jodrell Bank radio observatory
(\url{http://www.jb.man.ac.uk/~pulsar/crab.html}; \citealt{1993MNRAS.265.1003L}).

Aqueye and Iqueye have been realized with the support of the University of Padova, the Italian Ministry of Research and University MIUR, the Italian Institute of Astrophysics INAF, and the Fondazione Cariparo Padova.

L.~Z.~acknowledges financial support from the Italian Space Agency (ASI) and National Institute for Astrophysics (INAF) under agreements ASI-INAF I/037/12/0 and ASI-INAF n.2017-14-H.0 and from INAF ``Sostegno alla ricerca scientifica main streams dell'INAF'' Presidential Decree 43/2018.

We thank the developers of \texttt{astropy} \citep{2013A&A...558A..33A,2018AJ....156..123A}, \texttt{numpy} \citep{2020Natur.585..357H}, \texttt{scipy} \citep{2020NatMe..17..261V}, \texttt{matplotlib} \citep{2007CSE.....9...90H}, and \texttt{stingray} \citep{2019ApJ...881...39H}.

T.~Cassanelli wishes to thank Paul Scholz and Jing Luo for their helpful discussions about epoch folding, radio pulsar folding, and X-ray observations.
\end{acknowledgements}

\bibliographystyle{aa.bst}          % style aa.bst
\bibliography{wpca_folding.bib}  % bibliography file

\end{document}